\documentclass[preprint,12pt]{elsarticle}
\newcommand{\mrm}[1]{\mathrm{#1}}
\newcommand{\mbf}[1]{\mathbf{#1}}
\def\d{\mathrm{d}}

\def\QQ{\mathbf{Q}}
\def\pp{\mathbf{p}}
\def\qq{\mathbf{q}}

\def\kk{\mathbf{k}}

\def\rr{\mathbf{r}}

\usepackage{color} 
  
  \def\nuc#1#2{\relax\ifmmode{}^{#1}{\protect\text{#2}}\else${}^{#1}$#2\fi}
  \def\itnuc#1#2{\setbox\@tempboxa=\hbox{\scriptsize\it #1}
    \def\@tempa{{}^{\box\@tempboxa}\!\protect\text{\it #2}}\relax
    \ifmmode \@tempa \else $\@tempa$\fi}

\usepackage{amssymb}

\journal{Physics Letters B}

\begin{document}

\begin{frontmatter}


\title{Effective field theory for proton halo nuclei}


\author[chalmers]{Emil Ryberg}
\author[chalmers]{Christian Forss\'en}
\author[bonn,darmstadt,emmi]{H.-W. Hammer}
\author[chalmers,anl]{Lucas Platter\corref{lp}}
\cortext[lp]{lplatter@phy.anl.gov}

\address[chalmers]{Department of Fundamental Physics, Chalmers
  University of Technology,\\ SE-412 96 G\"oteborg, Sweden} 
\address[bonn]{Helmholtz-Institut f\"ur Strahlen- und Kernphysik,
  Universit\"at Bonn, 53115, Bonn, Germany}
\address[darmstadt]{Institut f\"ur Kernphysik, 
Technische Universit\"at Darmstadt, 64289 Darmstadt, Germany}
\address[emmi]{ExtreMe Matter Institute EMMI, GSI Helmholtzzentrum f\"ur 
Schwerionenforschung, 64291 Darmstadt, Germany}
\address[anl]{Physics Division, Argonne National Laboratory, Argonne,
  Illinois 60439, USA}

\begin{abstract}
  We use halo effective field theory to analyze the universal features
  of proton halo nuclei bound due to a large S-wave scattering
  length. With a Lagrangian built from effective core and
  valence-proton fields, 
  we derive a leading-order expression for the charge form
  factor. Within the same framework we also calculate the radiative
  proton capture cross section. Our general results at leading order are applied
  to study the excited $1/2^+$ state of Fluorine-17, and we
  give results for the charge radius and the astrophysical
  S-factor.

\end{abstract}


\end{frontmatter}


\section{Introduction
  \label{sec:intro}} 
Exotic isotopes along the neutron and proton drip lines are important
for our understanding of the formation of elements and they constitute
tests of our understanding of nuclear structure. The proton- and
neutron-rich regimes in the chart of nuclei are therefore the focus of
existing and forthcoming experimental facilities around the
world~\cite{Thoennessen:2011ke}.  The emergence of new degrees of
freedom is one important feature of these systems; exemplified, e.g., by
the discovery of several nuclear halo states along the drip
lines~\cite{2004PhR...389....1J,Jensen:2004eb,Riisager:2013bz}. Halo
states in nuclei are characterized by a tightly bound core with weakly
attached valence nucleon(s). Universal structures of such states can be
considered a consequence of quantum tunneling, where tightly-bound
clusters of nucleons behave coherently at low energies and the dynamics
is dominated by relative motion at distances beyond the region of the
short-range interaction.  In the absence of the Coulomb interaction, it
is known that halo nuclei bound due to a large positive S-wave
scattering length will show universal
features~\cite{2006PhR...428..259B,Hammer:2011kg}.  
In the case of proton halo nuclei, however, the Coulomb interaction introduces 
an additional momentum scale $k_{\mrm C}$, which is proportional to the charge 
of the core and the reduced mass of the halo system.
The low-energy properties of proton halos strongly depend on
$k_{\mrm C}$.

Halo effective field theory (EFT) is the ideal tool to analyze the
features of halo states with a minimal set of assumptions. It
describes these systems using their effective degrees of freedom,
i.e. core and valence nucleons, and interactions that are dictated by
low-energy constants \cite{Bertulani:2002sz,Bedaque:2003wa}.
For S-wave proton halo systems there will be a
single unknown coupling constant at leading order, and this parameter
can be determined from the experimental scattering length, or the
one-proton separation energy.
Obviously, halo EFT is not intended to compete with \emph{ab initio}
calculations that, if applicable, would aim to predict low-energy 
observables from computations starting with a microscopic description of
the many-body system. Instead, halo EFT is complementary to such
approaches as it provides a low-energy description of these systems
in terms of effective degrees of freedom.
This reduces the complexity of the problem significantly. By
construction, it can also aid to elucidate the relationship between different
low-energy observables.

Furthermore, halo EFT is built on fields for clusters, which makes it
related to phenomenological few-body cluster
models~\cite{Jensen:2004eb}. The latter have often been used
successfully for confrontation with data for specific processes
involving halo nuclei. A relevant example in the current context is the
study of proton radiative capture into
low-lying states states of \nuc{17}{F}~\cite{Rolfs:1973cj}. 
A general discussion of electromagnetic reactions of
proton halos in a cluster approach was given in \cite{Typel:2004us}.
The emphasis of an EFT, however, is the
systematic expansion of the most general interactions and, as a
consequence, the ability to estimate errors and to improve predictions
order by order. The structure and reactions of one- and two-neutron 
halos have been studied in halo EFT
over the last years (see, e.g., Refs.~\cite{Canham:2008jd,Hammer:2011ye,
Rupak:2011nk,Rupak:2012cr,Rotureau:2012yu,
Hagen:2013xga,Acharya:2013aea}). However,
concerning charged systems only unbound states such as $\alpha
\alpha$~\cite{Higa:2008dn} and $\alpha p$~\cite{Higa:2010di} have been
treated in halo EFT.

In this letter, we apply halo EFT for the first time to one-proton
halo nuclei. We restrict ourselves to leading order calculations of
systems that are bound due to a large S-wave scattering length between
the core and the proton. The manuscript is organized as follows: In
Sec.~\ref{sec:theory}, we introduce the halo EFT and discuss how
Coulomb interactions are treated within this framework. In the
following section, we present our results and calculate, in particular,
the charge form factor and charge radius at leading
order. Furthermore, we derive expressions for the radiative capture
cross section. We apply our general formulae to the excited $1/2^+$
state of \nuc{17}{F} and compare our numerical results with existing
data for this system.  We conclude with an outlook and a discussion on the
importance of higher-order corrections.
\section{Theory
\label{sec:theory}}
In halo EFT, the core and the valence nucleons are taken as the
degrees of freedom. For a one-proton halo system, the
Lagrangian is given by
\begin{equation}
\label{eq:lagrangian}
\mathcal{L}=\sum_{k=0,1}\psi_k^\dagger\left(i\mrm{D}_0+\frac{\mbf{D}^2}{2m_k}\right)\psi_k
-C_0\psi_0^\dagger\psi_1^\dagger
\psi_1\psi_0
+\ldots
\end{equation}
Here $\psi_0$ denotes the proton field with mass $m_0$ and $\psi_1$
the core field with mass $m_1$, $C_0$ denotes the leading order (LO)
coupling constant, and the dots denote derivative operators that 
facilitate the calculation
of higher order corrections. The covariant derivative is defined as
$\mrm{D}_\mu:=\partial_\mu+i e \hat{\mrm{Q}} A_\mu~$, 
where $\hat{\mrm{Q}}$ is the charge operator. The resulting
one-particle propagator is given by
\begin{equation}
  \label{eq:protonprop}
  iS_{k}(p_0,\pp)=i\left[p_0-\frac{\pp^2}{2m_k}+i\varepsilon\right]^{-1}~.
\end{equation}
For convenience, we will also define the proton-core two-particle
propagator
\begin{equation}
  iS_{\rm tot}(p_0,\pp)=i\left[ p_0-\frac{\pp^2}{2 m_{\rm R}}+i\varepsilon\right]^{-1}~,
\end{equation}
where $m_{\rm R}$ denotes the reduced mass of the proton-core system.
We include the Coulomb interaction through the full Coulomb Green's
function 
\begin{equation}
\langle \kk | G_\mrm{C}(E)|\pp\rangle = -S_{\rm
  tot}(E,\kk)\chi(\kk,\pp;E) S_{\rm tot}(E,\pp)~,
\label{eqCGFGamma}
\end{equation}
\begin{figure}[t]
\centerline{
\includegraphics*[width=8cm,clip=true]{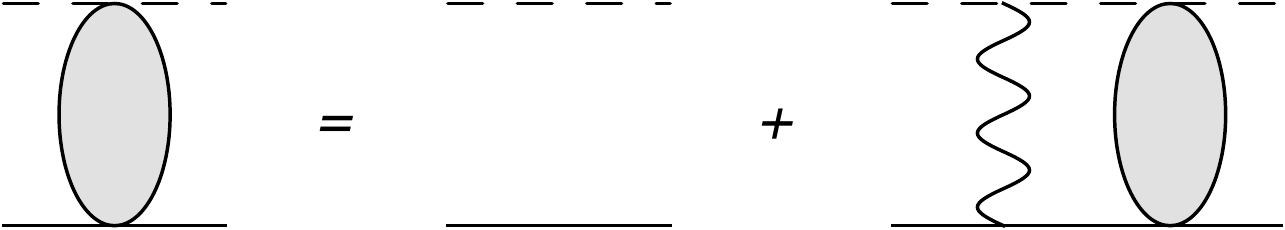}}
\caption{The integral equation for the four-point function 
$\chi(\kk_1,\kk_2)~$. The dashed line denotes a core propagator, 
the solid line a proton propagator and the wave line denotes the 
exchange of a Coulomb photon.}
\label{fig:fourpointgamma}
\end{figure}
where $\chi$ is the Coulomb four-point function defined recursively in
Fig.~\ref{fig:fourpointgamma}.  To distinguish coordinate space from
momentum space states we will denote the former with round brackets,
i.e. $(\rr|$. In coordinate space, the Coulomb Green's function can be
expressed via its spectral representation
\begin{equation}
(\rr|G_\mrm{C}(E)|\rr')=\int\frac{\d^3p}{(2\pi)^3}\frac{\psi_\pp(\rr)\psi^*_\pp(\rr')}{E-\pp^2/(2m_\mrm{R})+i\varepsilon}~,
\label{eq:CGFSpectral}
\end{equation}
where we define the Coulomb wave function through its partial
wave expansion
\begin{equation}
\psi_\pp(\rr)=\sum_{l=0}^{\infty}(2l+1)i^l\exp{(i\sigma_l)\frac{F_l(\eta,\rho)}{\rho}P_l(\hat{\pp}\cdot\hat{\rr})}~.
\end{equation}
Here we have defined $\rho=pr$ and $\eta= k_\mrm{C}/p~,$ with the
Coulomb momentum $ k_\mrm{C}=Z_\mrm{c}\alpha m_\mrm{R}~,$ and also the
pure Coulomb phase shift $\sigma_l=\arg{\Gamma(l+1+i\eta)}~.$ 
For the Coulomb functions $F_l$ and $G_l$, we use the 
conventions of Ref.~\cite{Koenig:2012bv}.
The
regular Coulomb function $F_l$ can be expressed in terms of the Whittaker M-function
according to
\begin{equation}
F_l(\eta,\rho)=A_l(\eta)M_{i\eta,l+1/2}(2i\rho)~,
\end{equation}
with the $A_l$ defined as
\begin{equation}
A_l(\eta)=\frac{|\Gamma{(l+1+i\eta)}|\exp{\left[-\pi\eta/2-i(l+1)\pi/2\right]}}{2(2l+1)!}~.
\end{equation}
We shall also need the irregular Coulomb wave function, $G_l$, which is given by
 \begin{equation}
G_l(\eta,\rho)=iF_l(\eta,\rho)+B_l(\eta)W_{i\eta,l+1/2}(2i\rho)~,
\end{equation}
where $W$ is the Whittaker W-function and the coefficient $B_l$ is defined as
\begin{equation}
B_l(\eta)=\frac{\exp{(\pi\eta/2+il\pi/2)}}{\arg{\Gamma{(l+1+i\eta)}}}~.
\end{equation}

\begin{figure}[t]
\centerline{
\includegraphics*[width=4cm,clip=true]{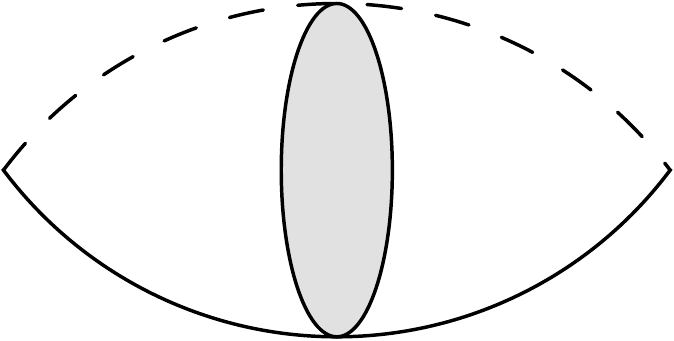}}
\caption{The irreducible self-energy at leading order. The solid
  (dashed) line denotes the proton (core) propagator. The shaded blob
  denotes the Coulomb Green's function.}
\label{fig:irredbubble}
\end{figure}

To obtain the fully-dressed two-particle propagator, that includes
strong and Coulomb interactions, we calculate the irreducible
self-energy shown in Fig.~\ref{fig:irredbubble}.
\begin{eqnarray}
i\Sigma(E)
&=&-i\int\frac{\d^3k_1\d^3k_2}{(2\pi)^6} S_{\rm
  tot}(E,\kk_1)\chi(\kk_1,\kk_2) S_{\rm tot}(E,\kk_2)
\nonumber\\
&=&i(0 | G_\mrm{C}(E)|0)~.
\end{eqnarray}

The expression above is known and is given by
\begin{equation}
(0 | G_\mrm{C}(E)|0)=-2
m_R\int\frac{\d^3q}{(2\pi)^3}\frac{\psi_\qq(0)\psi^*_\qq(0)}{q^2-2 m_R E-i\varepsilon}~.
\label{eqSigmaJ0}
\end{equation}
This integral was solved in Ref.~\cite{Kong:1999sf}, using dimensional
regularization in the power divergence subtraction (PDS) scheme, as
\begin{equation}
\Sigma(E)=-\frac{k_\mathrm{C} m_\mathrm{R}}{\pi}H(\eta)+\Sigma^\mrm{div}~,
\label{eq:SigmaSolved}
\end{equation}
with a divergent part $\Sigma^\mrm{div}$
\begin{equation}
\label{eq:Jdivergent}
\Sigma^\mrm{div}=\frac{k_\mathrm{C}
  m_\mathrm{R}}{\pi}\left[\frac{1}{3-d}+\log\left({\frac{\sqrt{\pi}\mu}{2 k_\mathrm{C}}}\right)
+1-\frac{3C_\mrm{E}}{2}\right]-\frac{m_\mathrm{R}\mu}{2\pi}~,
\end{equation}
where $d$ is the space dimension, 
$C_\mrm{E}$ is the Euler constant, and $\mu$ is the PDS regulator. 
The function $H$ is defined as
\begin{equation}
H(\eta)=\psi(i\eta)+\frac{1}{2i\eta}-\log{(i\eta)}~,
\end{equation}
with $\psi$ being the polygamma function. Note that the divergent part
in Eq.~(\ref{eq:Jdivergent}) is energy independent. This will become
important later when the derivative of $\Sigma$, with respect to the
energy, will be required.

The coupling constant $C_0$ can be determined by matching
to a two-body observable, such as 
the Coulomb corrected proton-core scattering length \cite{Kong:1999sf}:
\begin{equation}
\frac{1}{a_\mathrm{C}}=\frac{2\pi}{m_\mathrm{R}}\left(\frac{1}{C_0}-\Sigma^\mrm{div}\right)~.
\end{equation}
Since we stay at leading order, however, an explicit expression for 
$C_0$ will not be required for the calculation of electromagnetic
observables in the next section.

\section{Results
\label{sec:results}}
\subsection{Charge Form Factor}
\begin{figure}[t]
\centerline{\includegraphics*[height=3cm,angle=0,clip=true]{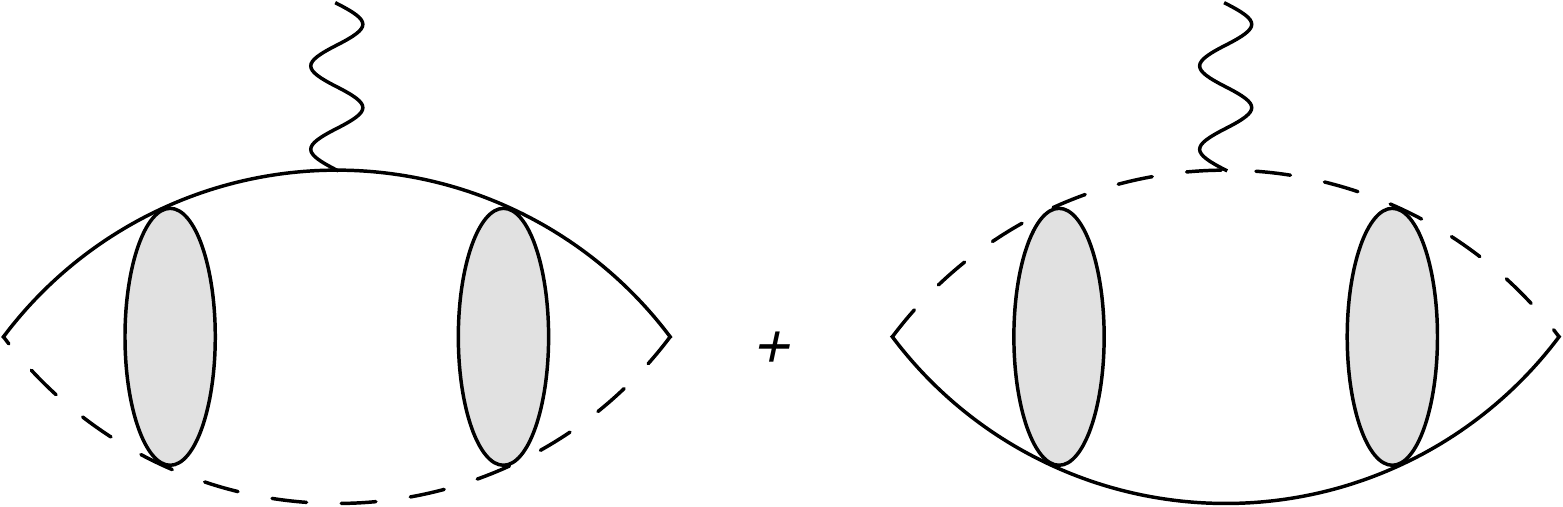}}
\caption{The irreducible three-point function $\Gamma^0~$.}
\label{figLOGamma}
\end{figure}
In our calculation of the charge form factor, we follow the derivation of the
deuteron form factor presented in Ref.~\cite{KSW99}. The form factor
is obtained by calculating the matrix element
\begin{equation}
  \langle \pp'|J^0_\mrm{EM}|\pp \rangle= e(Z_c+1) F_\mrm{C}( \QQ^2)~,
\label{eqDefFormFactor}
\end{equation}
for momentum transfer $\QQ=\pp'-\pp$ in the Breit frame, where no 
energy is transferred by the photon.  It
was shown in Ref.~\cite{KSW99} that this matrix element can be
expressed as\footnote{The reason for the additional factor of $-i$ in our Eq. (\ref{eq:ff1}) compared to Eq. (A10) of \cite{KSW99} is that we have an extra $-i$ in our definition of the irreducible self-energy.}
\begin{eqnarray}
  \label{eq:ff1}
  \langle \pp' | J^0_\mrm{EM}|\pp\rangle
&=&\frac{\Gamma^0(\QQ)}{\Sigma'(-B)}~,
\end{eqnarray}
where $\Gamma^0$ denotes the irreducible three-point function shown in
Fig.~\ref{figLOGamma}, and $\Sigma'(-B)$ is the derivative of the
self-energy with respect to the total energy evaluated at the energy
$E=-B~$, where $B$ is the proton separation energy or core-proton binding energy.

With the proton-core mass ratio $f=m_0/m_1~$, the three-point 
function $\Gamma^0$ 
is given by
\begin{eqnarray}
\nonumber
\Gamma^0(\QQ)&=&-e Z_\mrm{c}\int\d^3r\exp{(if\QQ\cdot\rr)}
|(0|G_\mrm{C}(-B)|\rr)|^2
\\
&&+\Big[(f\to1-f),~(Z_\mrm{c}\to 1)\Big]~,
\label{eqLOGamma}
\end{eqnarray}
and the derivative of the self-energy can be written as
\begin{equation}
  \Sigma'(-B)=
-\int\frac{\d^3q}{(2\pi)^3}\frac{\psi_\qq(0)\psi^*_\qq(0)}{(B+\qq^2/(2
  m_R))^2}~.
\label{eq:dSigmaE}
\end{equation}
Evaluating $\Gamma^0$ at zero momentum transfer, by using
Eq.~(\ref{eq:CGFSpectral}) and orthonormality of the wavefunctions,
and comparing with Eq.~(\ref{eq:dSigmaE}) shows that the charge form
factor is properly normalized to one in this limit.

We find that Eq.~(\ref{eqLOGamma}) can be simplified by writing the Coulomb
Green's function for negative energy using the Whittaker W-function.
This is achieved by demanding proper asymptotics and using that only
the S-wave part can contribute to propagation to zero separation, that is
\begin{eqnarray}
(0|G_\mrm{C}(-B)|\rr)&=&\lim_{\rho'\to0}\left(-i\frac{m_\mathrm{R}\gamma_0}{2\pi}\frac{F_0(\eta,\rho')\left[iF_0(\eta,\rho)+G_0(\eta,\rho)\right]}{\rho'\rho}\right)
\nonumber\\
&=&i\frac{m_\mrm{R}\Gamma(1+k_\mrm{C}/\gamma_0)}{2\pi}\frac{W_{-k_\mrm{C}/\gamma_0,1/2}(2\gamma_0r)}{r}~,
\label{eq:SimplCoulGreen}
\end{eqnarray}
where we have introduced the binding momentum
$\gamma_0=\sqrt{2m_\mathrm{R}B}$. The resulting integral is then 
\begin{eqnarray}
\Gamma^0(\QQ)&=&-\frac{e m_\mrm{R}^2\Gamma(1+k_\mrm{C}/\gamma_0)^2}{\pi}\int\hbox{d} r\left[Z_\mrm{c} j_0\Big(fQr\Big)+j_0\Big((1-f)Qr\Big)\right]\nonumber\\
&&\times W_{-k_\mrm{C}/\gamma_0,1/2}(2\gamma_0 r)^2~,
\label{eqBintegralnprime}
\end{eqnarray}
where $j_l$ are the spherical Bessel functions.
Once the parameters of the proton halo system are fixed, the equation
\begin{equation}
F_\mrm{C}(\QQ^2)=\frac{\Gamma^0(\QQ)}{e(Z_c+1)\Sigma'(-B)}
\end{equation}
is used to calculate the charge form factor and the corresponding charge
radius numerically. We have calculated these quantities for the
excited $1/2^+$ state of \nuc{17}{F}, which has a proton separation energy
of $B=104.94~(35)~\mathrm{keV}~$\cite{Tilley:1993}. Note that the
proton separation energy is the only non-trivial experimental input at
LO.

The charge form factor is related to the charge radius via the
expansion
\begin{eqnarray}
  F_\mrm{C}(\QQ^2)=1-\frac{\langle r_\mrm{C}^2\rangle_\mathrm{rel}}{6} \QQ^2+\ldots~,
\end{eqnarray}
and we find for the charge radius squared
\begin{equation}
  \label{eq:diffchargeradius}
  \langle r_\mrm{C}^2\rangle_\mathrm{rel}=(0.59~\mrm{fm})^2~.
\end{equation}
Since the proton and core are treated as structureless fields in halo EFT,
this quantity corresponds to the charge radius difference according to 
\begin{equation}
  \label{eq:chargeradius}
  \langle r_\mrm{C}^2\rangle_{{}^{17}\mathrm{F}^*}=\frac{Z_{\mrm c}}{Z_{\mrm c}+1}\langle r_\mrm{C}^2\rangle_{{}^{16}\mathrm{O}}
+\frac{1}{Z_{\mrm c}+1}\langle r_\mrm{C}^2\rangle_{\mathrm{p}}
+\langle r_\mrm{C}^2\rangle_\mathrm{rel}~,
\end{equation}
where $\langle r_\mrm{C}^2\rangle_{\mathrm{X}}$ is the charge radius
squared corresponding to the
particle $\mathrm{X}= {}^{17}\mathrm{F}^*$, \nuc{16}{O}, $\mathrm{p}$. 
Analog to the deuteron case, the charge radii of proton and \nuc{16}{O} 
enter at higher orders in the calculation via counter terms.

The error of the EFT can be estimated by comparing the momentum scale
$k_\mathrm{lo}\sim\gamma_0$ of the halo with the break-down scale
$k_\mathrm{hi}$ of the EFT. The latter is given by the closest interfering
state. For the \nuc{17}{F} halo system the break-down scale is given
by the bound state, $E_0=495.33~(10)~\mathrm{keV}~$ below the $1/2^+$
state \cite{Tilley:1993}. Thus, the expected LO error is
$\gamma_0/\sqrt{2m_\mathrm{R}E_0}\approx 1/2$ for the halo state in
\nuc{17}{F}$^*$, which is comparable to the LO error of a pionless EFT
calculation for the two-nucleon system.

\subsection{Radiative Capture}
\begin{figure}[t]
\centerline{\includegraphics*[height=3cm,angle=0,clip=true]{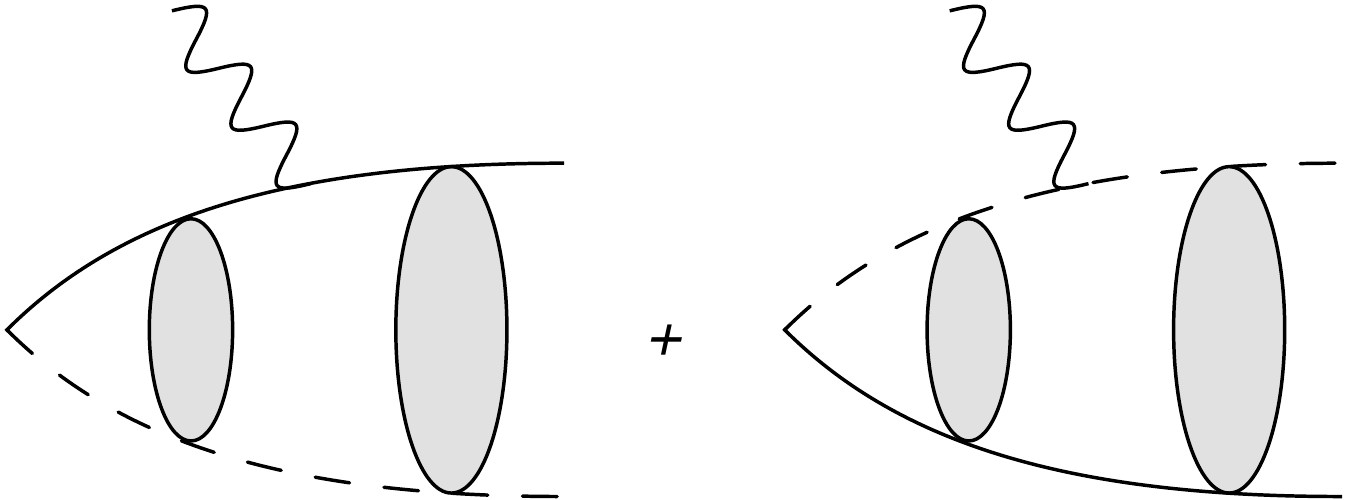}}
\caption{The radiative proton capture diagrams.}
\label{fig:Capture}
\end{figure}
Our approach can easily be applied to low-energy radiative capture.
The differential cross section for this reaction is
\begin{equation}
\frac{\d\sigma}{\d\Omega}=\frac{m_\mrm{R}\omega}{8\pi^2p}\sum_{i=1}^2\left|\epsilon_i\cdot\frac{\mathcal{A}}{\sqrt{\Sigma'(-B)}}\right|^2~,
\end{equation}
where $\mathcal{A}$ is the vector amplitude for the sum of the
diagrams shown in Fig.~\ref{fig:Capture}, where a proton is captured
by a core while a real photon is emitted. The relative momentum of the
proton-core system is $\pp$ and the four-momentum of the photon is
$(\omega,~\omega\hat{z})~$, with associated polarization vectors
$\epsilon_1=\hat{x}$ and $\epsilon_2=\hat{y}$. The factor
$1/\sqrt{\Sigma'(-B)}$ is the wavefunction renormalization, or LSZ
reduction factor.

The vector amplitude $\mathcal{A}$ can be expressed as the integral
\begin{eqnarray}
\mathcal{A}&=&\frac{eZ_\mrm{c}f}{m_\mrm{R}}\int\d^3r\,(0|G_\mrm{C}(-B)|\rr)\exp{(-if\omega r \cos{\theta})}\Big(\nabla\psi_\pp(\rr)\Big)\nonumber\\
&&+\Big[(f\to 1-f),~(Z_\mrm{c}\to 1)\Big]~,
\end{eqnarray}
where the $\nabla$ has emerged from the Feynman rule of the vector
photon coupling and acts on the Coulomb wavefunction due to a partial
integration. Evaluating the angular integrals and multiplying with the
polarization vector, the integral is simplified to
\begin{eqnarray}
\sum_{i=1}^2\left|\epsilon_i\cdot\mathcal{A}\right|^2&=&
\Bigg|-i\sin{\theta}(\cos{\phi}
+\sin{\phi})\frac{4\pi
  eZ_\mrm{c}f\exp{(i\sigma_1)}}{m_\mrm{R}p}\nonumber\\
&&\times\int\d r(0|G_\mrm{C}(-B)|\rr)j_0(f\omega
r)\frac{\partial}{\partial r}
\Big[rF_1(k_\mrm{C}/p,pr)\Big]\nonumber\\
&&+\Big[(f\to 1-f),~(Z_\mrm{c}\to 1)\Big]\Bigg|^2~,
\label{eq:CaptureIntSimpl}
\end{eqnarray}
where the angles $\theta$ and $\phi$ will be integrated over to give the total cross section.
For a given physical system, we can solve the integral in
Eq.~(\ref{eq:CaptureIntSimpl}) numerically using
Eq.~(\ref{eq:SimplCoulGreen}) for the Coulomb Green's function.

\begin{figure}[t]
\centerline{\includegraphics*[width=10cm,angle=0,clip=true]{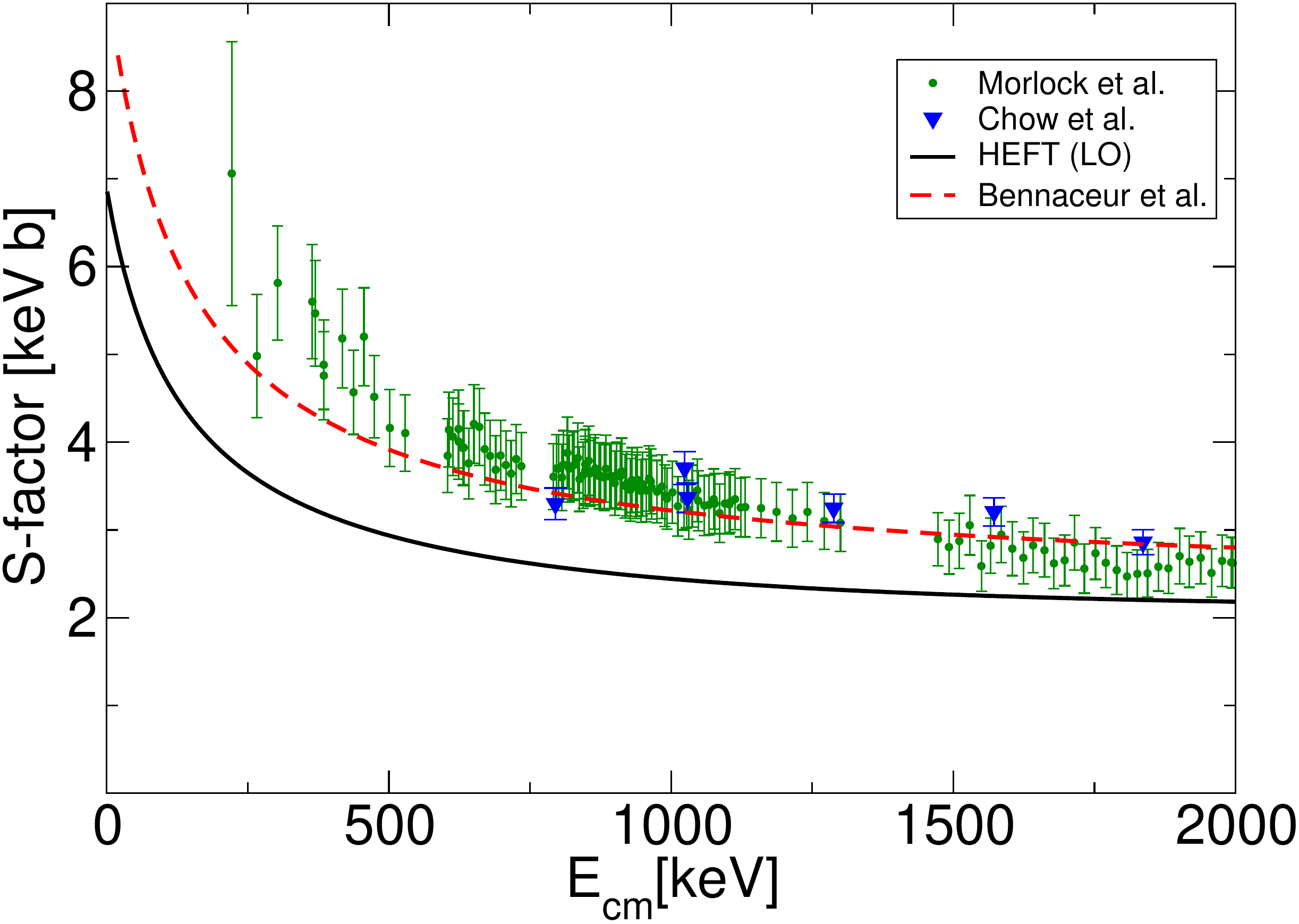}}
\caption{The LO halo EFT result for the astrophysical S-factor for
  \nuc{16}{O}(p,$\gamma$)\nuc{17}{F}$^{*}(1/2^+)$ is presented by the solid black line. The theoretical
  result is compared with the data by Chow {\it et al.}
  \cite{Chow:1975} and Morlock {\it et al.}
  \cite{Morlock:1997jh,Iliadis:2008id} shown by blue
  triangles and green dots, respectively. The calculation by Bennaceur {\it et
    al.} \cite{Bennaceur:2000da} is shown by the dashed curve.}
\label{fig:SFactor}
\end{figure}
Radiative capture into low-lying states of \nuc{17}{F} has been
measured by Rolfs {\it et al.} \cite{Rolfs:1973cj}, Chow {\it et al.}
\cite{Chow:1975} and by Morlock {\it et al.}
\cite{Morlock:1997jh,Iliadis:2008id}.
In Fig.~\ref{fig:SFactor} we show the astrophysical S-factor, defined as
\begin{equation}
\label{eq:Sfactor}
S(E)=E\exp{(2\pi\eta)}\sigma_\mrm{tot}~.
\end{equation}
The figure shows the halo EFT results of our LO calculation compared
to experimental data for capture into the $1/2^+$ excited state and a
phenomenological calculation using the shell model embedded in the
continuum. At threshold, we find that $S(0)\approx7~\mrm{keV~b}$. Our
LO results are slightly low, but consistent with the experimental
data, within the expected 50\% error. We anticipate that the
next-to-leading order correction will increase the radiative capture
cross section through the appearance of a finite effective range at
this order. It can also be noted that the results agree qualitatively
with the predictions obtained in the shell model embedded in the
continuum~\cite{Bennaceur:2000da}.
%
\section{Conclusions}
\label{sec:conclusions}
In this work, we have shown that Coulomb effects can be included in
halo EFT, and that thereby static and dynamical observables of proton
halo nuclei become accessible. We have calculated the charge radius
and the radiative proton capture cross section of S-wave proton halo
nuclei at leading order in halo EFT. Our results can be applied to any
one-proton halo system whose interaction is dominated by S-waves. In
particular, the excited $1/2^+$ state in \nuc{17}{F} is known to have
a large S-wave component. We have calculated the charge radius for
this system. While this observable is not yet experimentally
accessible, this result provides a prediction for \emph{ab initio}
calculations using modern nucleon-nucleon interactions.

In addition, we have compared our results for radiative capture into
the excited $1/2^+$ state of \nuc{17}{F} with experimental data and
found good agreement within the expected error. Furthermore, we found
that halo EFT gives the same qualitative behavior for this observable
as previous calculations that have employed phenomenological models.

For a quantitative description of the experimental data, higher order
corrections are required.  In a future publication, we will address
how these corrections are included within halo EFT in the presence of
Coulomb interactions. The size of these contributions will strongly be
affected by the relative size of the effective range and the Coulomb
momentum $k_\mathrm{C}$, which provides an additional scale in systems
with Coulomb interactions. Our calculation is also a first step
towards a calculation of properties of \nuc{8}{B} within halo
EFT. This system requires the inclusion of two low-energy constants at
leading order since it interacts dominantly in the P-wave
\cite{Bertulani:2002sz}.

Finally, our approach might prove useful for heavier systems whose
static observables can be calculated using \emph{ab initio} approaches,
but for which continuum properties are not accessible within the same
framework due to the computational complexity. In this scenario,
\emph{ab initio} predictions of, e.g., the one-proton separation energy
could be used to fix the halo EFT parameters, which in turn could be
used to predict continuum observables such as the radiative capture
cross section. In the case of neutron halos,
such an approach was recently carried out to predict
novel features in the Calcium isotope chain using halo
EFT~\cite{2013arXiv1306.3661H}.

\section*{Acknowledgement}
We thank H. Esbensen and S. K\"onig for helpful discussions, P. Mohr
and K. Bennaceur for supplying relevant data. This work was supported
by the Swedish Research Council (dnr. 2010-4078), the European
Research Council under the European Community’s Seventh Framework
Programme (FP7/2007-2013) / ERC grant agreement no.~240603, the Office
of Nuclear Physics, U.S.~Department of Energy under Contract
nos. DE-AC02-06CH11357, by the DFG and the NSFC through the
Sino-German CRC 110, by the BMBF under contract 05P12PDFTE, and by the
Helmholtz Association under contract HA216/EMMI.


\end{document}